\documentclass[runningheads]{llncs}
\usepackage{graphicx}
\usepackage{todonotes}
\usepackage{float}

\usepackage{tikzsymbols}
\usepackage{textcomp}

\begin{document}
\title{FakeYou! - A Gamified Approach for Building and Evaluating Resilience Against Fake News}
\titlerunning{FakeYou!}
%
\author{Lena Clever\inst{1}\orcidID{0000-0003-0926-6326} \and
Dennis Assenmacher\inst{1}\orcidID{0000-0001-9219-1956} \and
Kilian M\"uller\inst{1}\orcidID{0000-0001-7887-0223} \and
Moritz Vinzent Seiler\inst{1}\orcidID{0000-0002-1750-9060} \and
Dennis M. Riehle\inst{1}\orcidID{0000-0002-5071-2589} \and
Mike Preuss\inst{2}\orcidID{0000-0003-4681-1346} \and
Christian Grimme\inst{1}\orcidID{0000-0002-8608-8773}
}
\authorrunning{L. Clever et al.}
%
\institute{University of M\"unster, Department of Information Systems, Leonardo-Campus 3, 48149 M\"unster, Germany \and
University of Leiden, Niels Bohrweg 1, 2333 CA Leiden, The Netherlands}
\maketitle              
\begin{abstract}
Nowadays fake news are heavily discussed in public and political debates. Even though the phenomenon of intended false information is rather old, misinformation reaches a new level with the rise of the internet and participatory platforms. Due to Facebook and Co., purposeful false information - often called fake news - can be easily spread by everyone. Because of a high data volatility and variety in content types (text, images,...) debunking of fake news is a complex challenge. This is especially true for automated approaches, which are prone to fail validating the veracity of the information. This work focuses on an a gamified approach to strengthen the resilience of consumers towards fake news. The game FakeYou motivates its players to critically analyze headlines regarding their trustworthiness. Further, the game follows a "learning by doing strategy": by generating own fake headlines, users should experience the concepts of convincing fake headline formulations.
We introduce the game itself, as well as the underlying technical infrastructure. A first evaluation study shows, that users tend to use specific stylistic devices to generate fake news. Further, the results indicate, that creating good fakes and identifying correct headlines are challenging and hard to learn.

\end{abstract}
\section{Introduction \& Motivation}
\label{sec:intro}

Besides text, images are a traditional and mighty vehicle to transport (wrong) information into peoples minds~\cite{Bannatyne2019} making them most attractive for the purpose of intended misinformation - also called \emph{fake news}. While some researchers report on images being of significant importance for reaching a wider audience~\cite{Gunawan2020}, others show that information transported through (fabricated) images can change or even manipulate memories of viewers~\cite{Wade2002,Nash2009}. This is supported by some cognitive factors which render mentally digested misinformation resistant to correction~\cite{Sacchi2007,Lewandowski2012}. Very recent evidence confirms that multimodal disinformation, i.e., disinformation comprising text- and image-based information is more credible than just textual information~\cite{Hameleers2020}.

\begin{figure}
    \centering
    \includegraphics[width=0.6\textwidth]{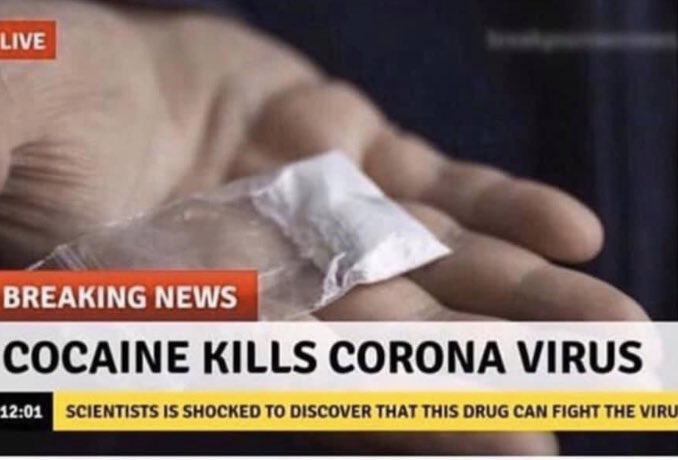}
    \caption{Fake image, that claims the corona virus breakout of 2019 in China could be cured by consuming cocaine. The image was debunked by the Mimikama project, \url{https://www.mimikama.at/allgemein/cocaine-kills-corona-virus/}.}
    \label{fig:fake_cocaine}
\end{figure}

Image fabrication has for long been a skill only feasible for experts but modern computers or simple-to-use online services enable virtually everybody to make up fake images. An example for the simplicity of image-based fake news generation is shown in Figure~\ref{fig:fake_cocaine}. Using the online service BreakYourOwnNews\footnote{\url{https://breakyourownnews.com/}}, a breaking news fake was produced that transported this 
misleading message.

With the rise of fake news~\cite{Bovet2019}, projects like Mimikama\footnote{\url{https://www.mimikama.at}} started to search for false messages in order to expose and debunk them. Much of their work focuses on images~\cite{Haigh2018}. Already before, research on Facebook~\cite{Friggeri2014} showed that especially image-based fakes cascade more deeply into social networks than correcting content. And of course, manual correction and research on each and every image is very time consuming making debunking permanently lagging behind. Also automation approaches for detecting fake news are not sufficient to solve the problem, as they are usually unable to validate textual as well as image-based content. Thus, current automation mainly addresses originality issues of images by trying to find whether an image was tempered or fabricated~\cite{Parikh2019,Dang2019}.

In this work, we focus on consumer resilience as another important building block of fighting fake news in practice. Instead of relying on external services like debunking and automated detection of manipulated images, we aim for a gamified approach 

\begin{enumerate}
    \item to sensitize social media consumers for the issue of multimodal (image- and text-based) fake news in general, 
    \item to demonstrate the individual challenges in evaluating presented information pieces in a restricted environment (like social media or news aggregator apps), and
    \item to enable consumers to experience and possibly develop techniques of generation for misleading information.
\end{enumerate}

All aspects are integrated into a single mobile application, in which users annotate original press photographs and images extracted from real news articles with fake text headlines. At the same time, users have to find the true headline in a multiple choice competition among fakes produced by other users. Both, successfully deceiving others and finding out the truth are rewarded.

As an intended side effect, this app is able to store any produced content and interaction data of users for further evaluation. As such, we provide this app as an education and evaluation platform for fostering and investigating resilience against fake news. The present work introduces the architecture and concept of this application and demonstrates a perspective for future research within a small case study with $N=53$ participants.

The work is structured as follows: Section~\ref{sec:background} gives a short overview on some current perspectives on fake news, the reception of misinformation and current research in the context of this work. Thereafter, Section~\ref{sec:architecture} provides a glimpse into   the game rules and concept, before Section~\ref{sec:game} introduced the aspects of the software's architecture and components. Section~\ref{sec:case_study} presents a case study on how user interaction and user generated content can be evaluated to learn about challenges in fake news detection and generation. The paper is concluded in Section~\ref{sec:discussion}.


\section{Related Work}
\label{sec:background}
The distribution and deceiving use of wrong or fabricated information is a rather old phenomenon~\cite{Bannatyne2019}. 
Historians in the pre-printing era used them as vehicles to influence the view of generations on a leader or emperors deeds~\cite{burkhardt2017} and information twisting certainly increased with the invention of printing techniques and the rise of mass media~\cite{Posetti2018}.
However, during the last decade and specifically with the emergence of the internet and social media, the term \emph{fake news} appeared in the public sphere.


In principle, the term still relates to false or fabricated information (misinformation) used for a specific, often disinformation-related, purpose. However, it is important to note that the understanding and usage of the term fake news have started to bifurcate. As Quandt et al. state, the term is now also used as ``a derogatory term denouncing media and journalism''~\cite{Quandt2019}. 

Apart from the increasingly blurry use of the term, three important factors changed compared to the pre-internet eras: (1) the fabrication of misinformation has become very simple due to computer and software technology advancements, (2) the global spreading of (mis)information is accessible to virtually everybody, and (3) information has become a commodity in modern life~\cite{Bannatyne2019}. This paves the ground for a massive increase of false information spread in social media, which is observable over the last years~\cite{Bannatyne2019}.

With the increasing relevance of intended misinformation, research focuses on different aspects of fake news definitions~\cite{Tandoc2018} and cognitive effects but recently also on means for suppression and debunking. Due to the existence of misinformation long before the term fake news was coined, research is far more advanced in the investigation of cognitive effects of false information to the human memory and capabilities to process corrections. Consequently, cognitive sciences are quite sure that misinformation transported by images is capable of changing memories of viewers~\cite{Wade2002,Nash2009}. At the same time, cognitive processes seem to fill gaps in consumer memories with fake information and support conclusion models that are rather immune against correction efforts~\cite{Sacchi2007,Lewandowski2012}. Additionally, there is some evidence that repeated exposure to rumors and misinformation strengthen the belief in them~\cite{Balmas2014,Berinsky2017}. Consequently, action as well as research on countering the effect of fake news addresses the exposure of consumers. While some favor fact checking~\cite{Haigh2018} and information correction~\cite{Lewandowski2012} as reaction to fake news, Barrera et al.~\cite{Barrera2020} find that fact checking alone is not sufficient to change peoples mind. A more technical approach is followed by those who try to use machine learning and image forensics techniques in order to detect fabricated images by learning manipulation patterns~\cite{Dang2019,Parikh2019}.

Both streams (understanding of fake effects and mechanisms as well as technological support) are also addressed in gamified research projects that integrate consumers of information. Rozenbeek et al.~\cite{Roozenbeek2019} design a browser-based serious game\footnote{\url{https://getbadnews.com/}} that demonstrates users how polarisation, emotions, conspiracy theory, trolling, and impersonation are used for fake news production and spread. They use the gaming data of about 15,000 participants to demonstrate that the game helps in increasing resilience of participants against fake news. However, the gameplay is rather sophisticated and based on a time consuming click-through simulated game flow, as well as on mostly text messages. 
With the intention of studying the influence of guidance in gameplay, Lutzke et al.~\cite{Lutzke2019} exposed participants -- one group with guidelines on how to deal with information, a control group without guidelines -- in an online experiment to fake news. The authors find, that guided participants had a reduced likelihood to share or like fake messages afterwards.
Katsaounidou et al.~\cite{Katsaounidou2019} provide the MAthE fake news game, a serious game that addresses verification and correction techniques/services. Therefore, the game provides a simulated search engine, reverse image search, an image verification assistant, and a debunking site. The authors find preliminary indications for raised awareness regarding authentication and verification tools.

However, each fake news game has a rather sophisticated gameplay and usually a strong educational focus on fake news production techniques or verification to direct player attention as well as learning processes. In this work, we try to combine both fake news production and evaluation in a very simple rule set and highly competitive gameplay to increase player dedication. Players are not guided through an educational program but should get aware of the simplicity of faking and the complexity of evaluating multimodal information in a restricted (app) environment indirectly by playing.

\section{Game Rule Set}
\label{sec:game}
In the following, we will briefly introduce the game FakeYou. The two main goals of a player in the game FakeYou are:

\begin{enumerate}
    \item Create a convincing fake headline for a given newspaper article image.

    \item Figure out the correct headline of this image, by choosing one of 3 candidates, where one headline is the original headline of the newspaper article, and the others are given by two opponents.
\end{enumerate}


\begin{figure}[htb]
\begin{minipage}{0.2\textwidth}
    \centering
    \includegraphics[width=\linewidth]{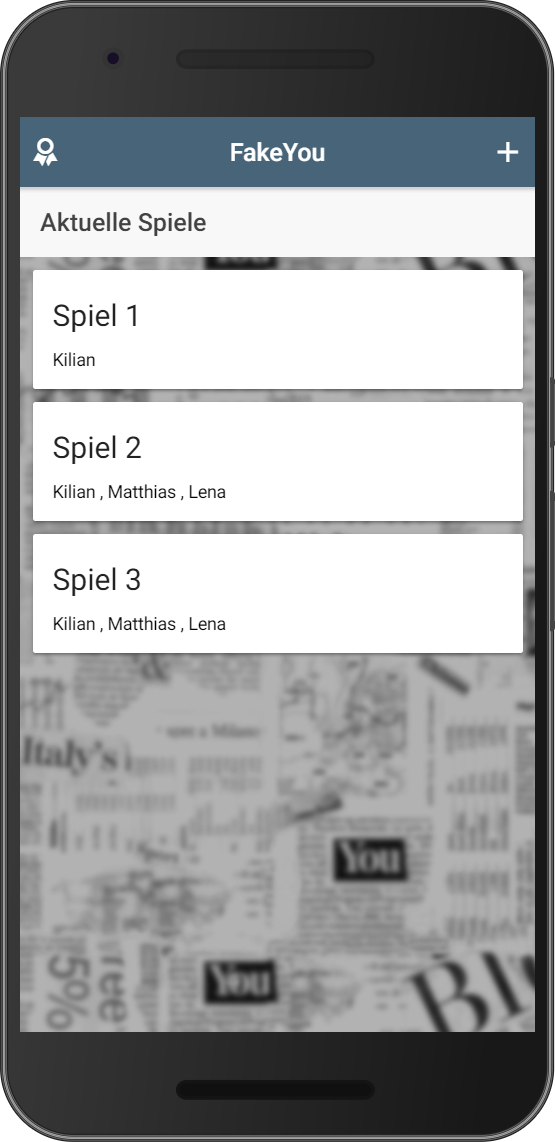}\\
    (a)
\end{minipage}
\hspace*{1.75cm}
\begin{minipage}{0.2\textwidth}
    \centering
    \includegraphics[width=\linewidth]{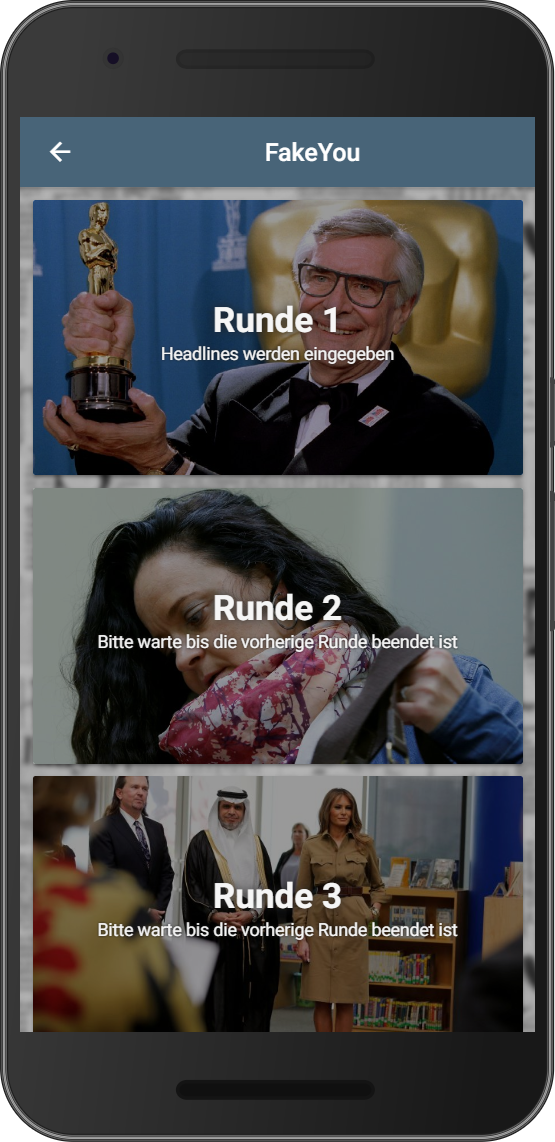}\\
    (b)
\end{minipage}
\hspace*{1.75cm}
\begin{minipage}{0.2\textwidth}
    \centering
    \includegraphics[width=\linewidth]{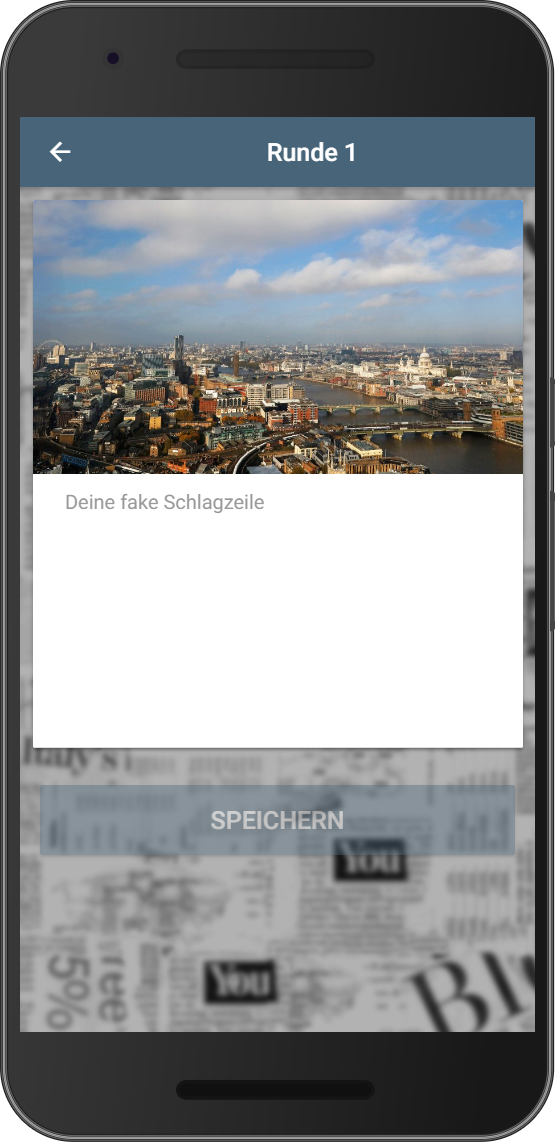}\\
    (c)
\end{minipage}
\vspace{0.3cm}

\begin{minipage}{0.2\textwidth}
\centering
\includegraphics[width=\linewidth]{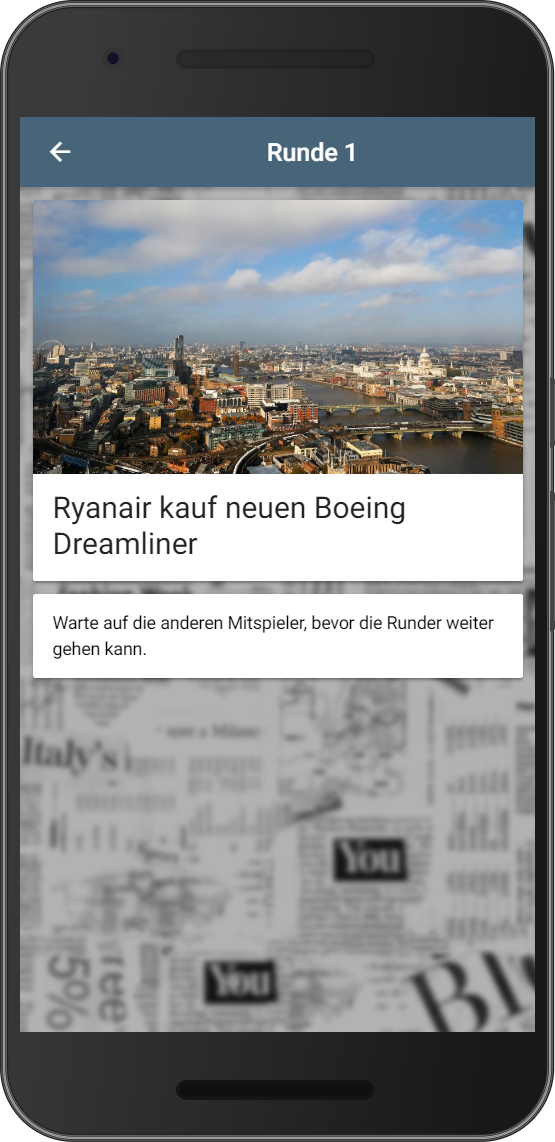}\\
(d)
\end{minipage}
\hspace*{1.75cm}
\begin{minipage}{0.2\textwidth}
\centering
\includegraphics[width=\linewidth]{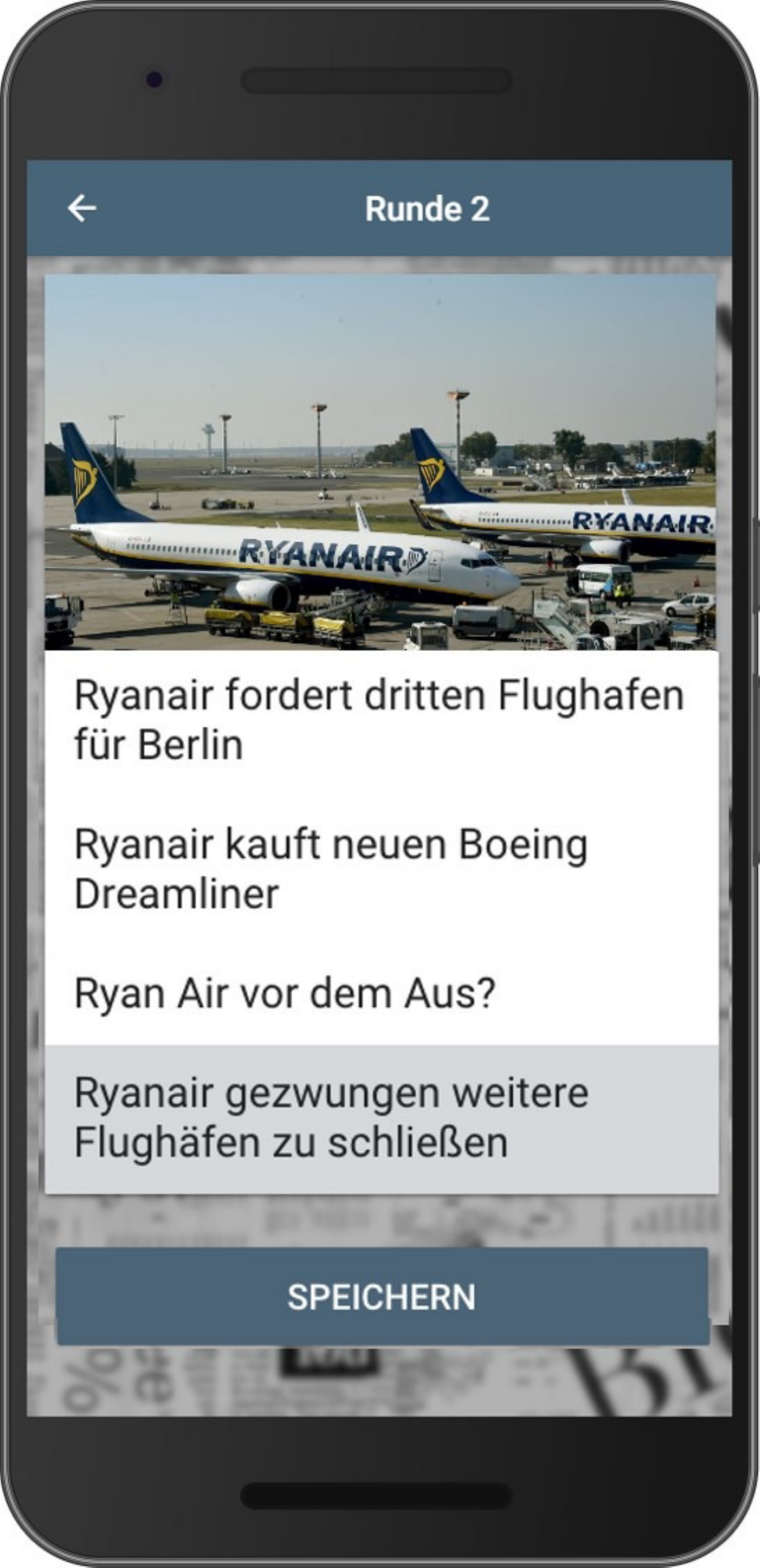}\\
(e)
\end{minipage}
\hspace*{1.75cm}
\begin{minipage}{0.2\textwidth}
\centering
\includegraphics[width=\linewidth]{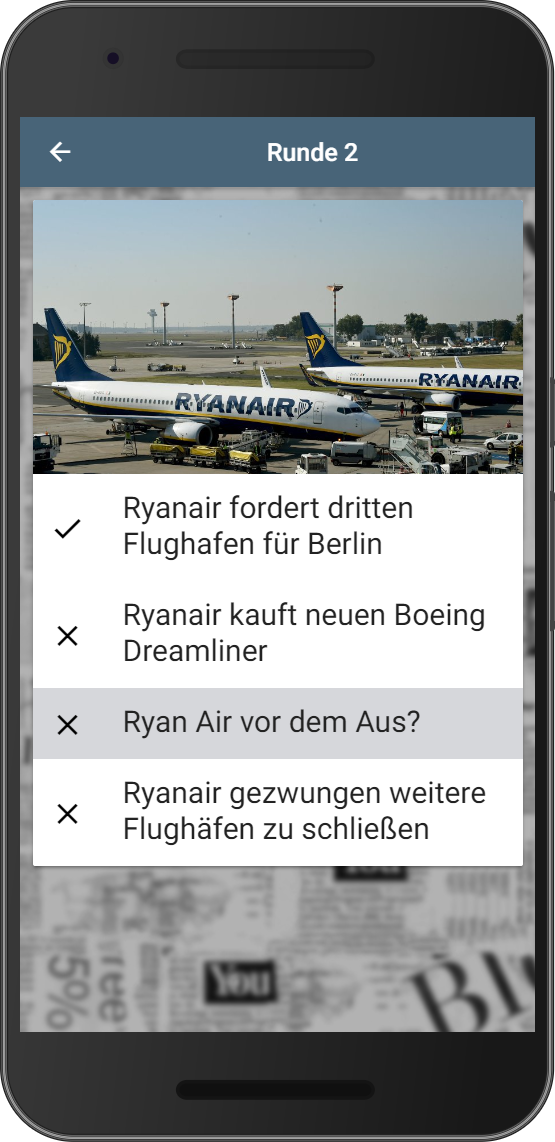}\\
(f)
\end{minipage}

\caption{The steps from (a) to (f) schematically describe the flow of the game and the ruleset of FakeYou.}
\label{fig:gameplay}
\end{figure}

After registration with an unique user name, the user accesses the game lobby (Figure~\ref{fig:gameplay}a). The game lobby consists of a list of started and finished games, as well as a button (+) in the right upper corner to start a new game. When a new game is started, the player has to wait until two other players opt to start a new game. As soon as three players are available, they are assigned to a new game and forwarded to the game page. 
Each game has three rounds. They are presented to the user next, see Figure~\ref{fig:gameplay}b. After selecting a round, the player can insert a suitable fake headline  for the given image (Figure~\ref{fig:gameplay}c). The goal is to create a fake headline, which is believed true by other players. When all three players inserted their headline, the round is forwarded to the evaluation step (Figure~\ref{fig:gameplay}d and e). 

Here, the correct headline has to be chosen out of three possible options (the two inserted headlines of the opponents and the original headline scarped with the picture). Picking the correct headline is scored by 2 points and fooling a player with a fake headline is scored by 3 points. 
After each player picked a headline, results are presented to the players (Figure~\ref{fig:gameplay}f). 


In the following section, a brief overview over the technical implementation and components of FakeYou is given.

\section{Architecture}
\label{sec:architecture}

\begin{figure}[htb]
    \centering
    \includegraphics[width=0.6\textwidth]{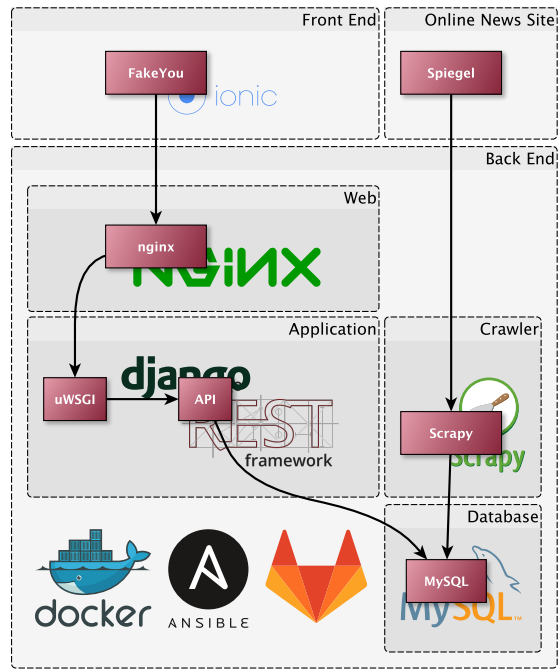}
    \caption{Architecture of Front and Back End}
    \label{F_7} 
\end{figure}

The general architecture of the game consists of a front end and a back end, as depicted in Figure~\ref{F_7}, where the back end is divided into different services. We will elaborate the main components of both, the front end and the back end of FakeYou, in the following sections.

\subsection{Front End}
FakeYou is designed as a mobile app allowing it to be played online on both personal computers as well as smartphones and tablets. Moreover, we implemented the game as a hybrid app in order to make it possible to play it with different operating systems such as \textit{Android} or \textit{iOS}, thus reaching a wider audience.

As depicted in the left upper corner of Figure~\ref{F_7}, the front end is developed with the help of the \textit{ionic} framework\footnote{See: \url{https://ionicframework.com/}}. \textit{Ionic} is an open source framework for the development of hybrid apps, which is built on Angular. It was used for the implementation of the user interface and played a fundamental role in the development of the eight pages, which are provided to the players in order to register, play the game, and view their relevant statistics. \textit{Angular 2}\footnote{See: \url{https://angular.io/}} is a \textit{TypeScript}-based, open source web application platform especially developed for front ends. Thus, it structures and connects the different views of the front end as well as offering multiple libraries for encryption and other features.

\subsection{Back End}
Apart from the front end and the third party information available on the internet, all information is stored within the back end as shown in the lower part of Figure \ref{F_7}. Information is accessed, encrypted, and transmitted via a \textit{nginx}\footnote{See: \url{https://www.nginx.com/}} web server and a \textit{django}\footnote{See: \url{https://www.djangoproject.com/}} REST framework. While \textit{nginx} acts as a proxy which facilitates the communication between the app and the back end, the django framework handles data access and the database via an API. There are two components to which the API is directly connected, the most important one being our database. As database we use \textit{MySQL}\footnote{See: \url{https://www.mysql.com/de/}}. The pictures required for FakeYou are stored on the hard disk, only storing the paths leading to the pictures in \textit{MySQL}. Apart from the pictures, all further important information required for FakeYou e.g. the user identification, scores, authentication tokens, and statistics are stored in \textit{MySQL}. Neither the app itself nor the web server has direct access to the database. Consequently, the database always delivers a complete and correct picture of all relevant data.

In order to fill the game with pictures and their corresponding headlines, we make use of a web crawler called \textit{Scrapy}\footnote{See: \url{https://scrapy.org/}}. With its help, we are able to store the connected URLs, headlines, publication dates, and languages from articles published on the crawled news websites in the database. The crawler automatically accesses the relevant news websites and retrieves and stores the headline links in specified time intervals, thus always providing new headlines as well as pictures. However, at the time we conducted our evaluation study, only the German version of \textit{Spiegel Online} could be crawled. Therefore, the only language available for our game was German. 


\section{Case Study}
\label{sec:case_study}
To get preliminary insights into the educational effects of our game and exemplary show interesting aspects that can be analyzed by using our tool, we conducted an evaluation case study with a small number of volunteers (mostly students and faculty members), who played the game and afterwards answered a questionnaire about their personal experience of the game. It should be emphasized that this rather small study with its exploratory analysis is only intended as a  showcase, or proof of concept, to motivate the diverse applications of our tool.   

\subsection{Study Setup and Data}
\begin{figure}
    \centering
    \includegraphics[width=\textwidth]{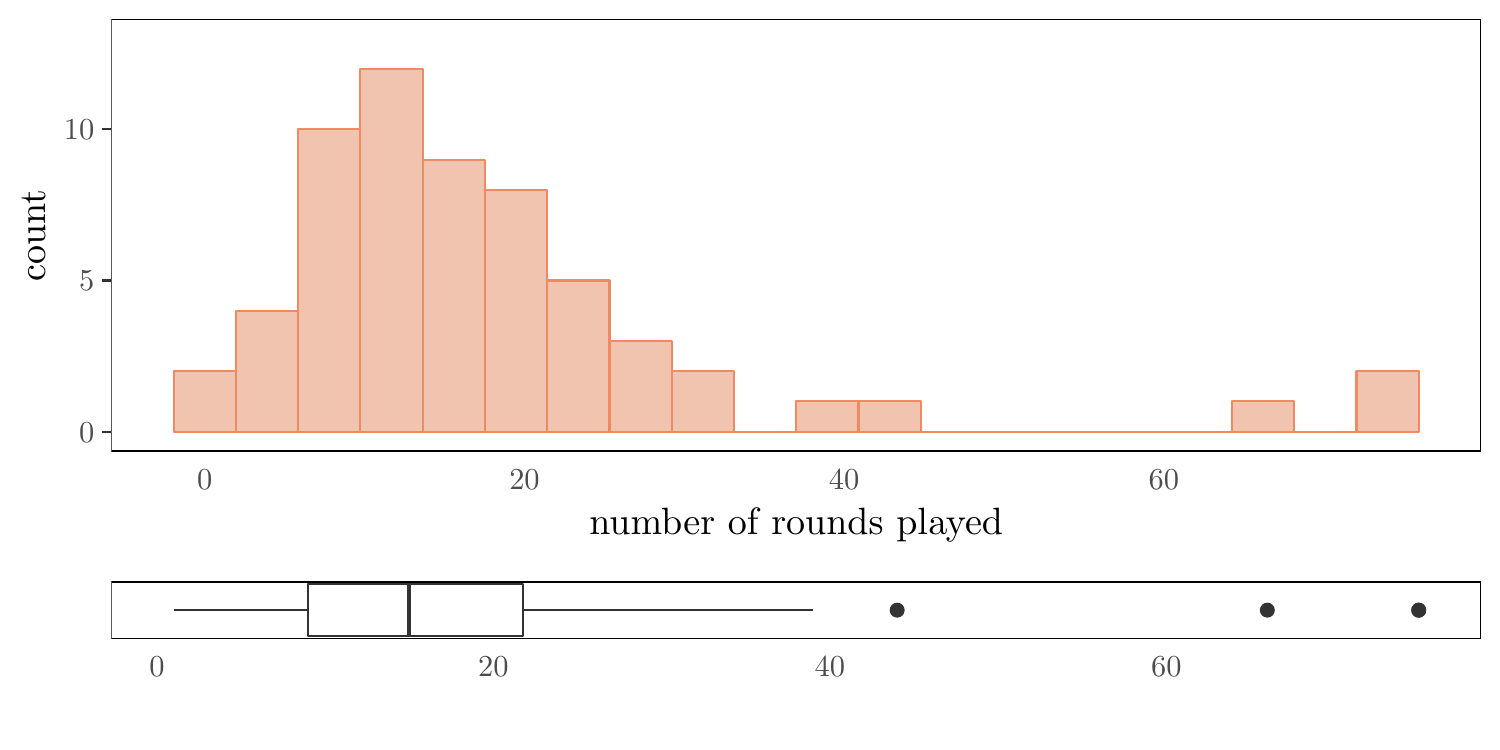}
    \caption{Deviation of number of rounds played during the evaluation study.}
    \label{fig:rounds}
\end{figure}

In total, 53 persons participated in the game ($75\%$ male, $25\%$ female). 
The gaming time varied between 30 minutes up to two hours. However, the number of games the players had played in these specified time intervals varied considerably from player to player. The amount of rounds played by every user during the case study is depicted in Figure~\ref{fig:rounds} and varied between 1 and 75. Fifty percent of the participants played 12 to 24 rounds which equals 4 to 8 games. There are only a few \textit{super users} who played fake you up to 75 rounds (25 games).

In total, 311 headlines were crawled from the \textit{Spiegel Online} website were used during the evaluation study and $1,080$ fake headlines were created by the players.
The data collection consisted of two parts. First, we invited the participants to play the game as often as they wanted within a time interval of seven hours. By this, we were able to collect data including the participants fake headlines, their opponents, the correct headlines they were able to detect, the headlines where they were fooled by other users, as well as whom and how often they were able to fool. Additionally, we gathered some metadata such as the number of games played by each user, the scores for every round and some further information like cancelled games.

After playing the game, we asked the participants to complete a questionnaire, which we conducted for two reasons: first, it was our intention to learn more about the players' gaming experience and the handling of the game. Besides, we asked them to provide us with suggestions regarding how we could further improve the app. Secondly, we collected additional relevant data for our analyses such as demographic data of the players (gender and age), their playing times, and how difficult it was for them to come up with fake news and to distinguish fake from real news. Of particular interest for our analyses were the answers concerning whether they were subjectively able to improve in playing the game over time.

\subsection{Ethics and Legal Aspects}
During the experiment no personal data has been collected or stored. Participants chose an artificial user nickname/alias to play the game. Therefore, no connection between the account and the actual legal person can be established. The same applies for the questionnaires. Further, it should be emphasized that the game is evaluated within an experimental setting. Images and crawled headlines from the news outlet were only accessible within the game environment during our experiment. To avoid copyright violations, the game and the image- and headline-database were only accessible in the evaluation environment.


\subsection{Analysis}
Within our analyses we tried to (a) identify specific patterns that are utilized by the users to create fake headlines (and do not occur within the original headlines) and (b) investigate whether we can identify some improvements in both, fake news creation and identification on an objective and subjective level. Based on our experiment, we therefore analyzed fake and original headlines in terms of word and character usage. Further, we elaborated the performance of players regarding their ability to fool their opponents and select the correct headline. Lastly, we evaluated, whether the players followed a learning curve during their game play. Additionally, we analyzed the questionnaires regarding the players perceived game experiences. 

\begin{figure}
    \centering
    \includegraphics[width=\textwidth]{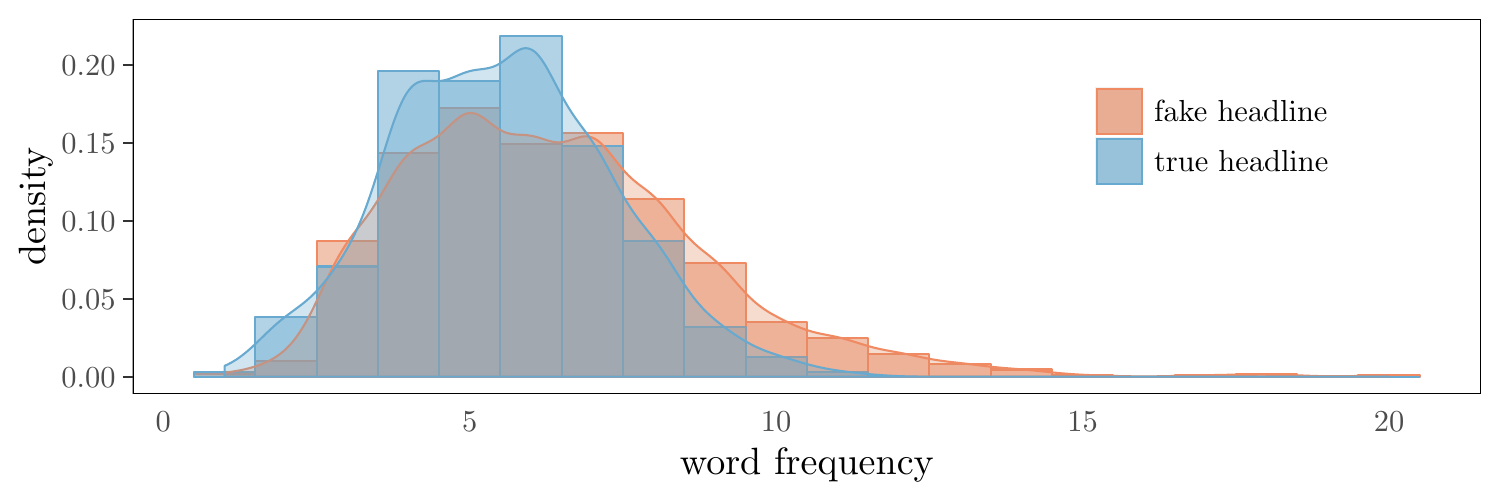}
    \caption{Word frequency density in fake and correct headlines.}
    \label{fig:word_hist}
\end{figure}

Figure~\ref{fig:word_hist} depicts the amount of words used in both the fake (orange) and the correct (blue) headlines. The amount of words used within a headline is stated on the x-axis, while the y-axis displays the density of both types of headlines. Both distributions are normalized due to the unequal number of fake and original headlines. The two distributions are significantly different according to a conducted Wilcoxon Rank-Sum Test ($p \leq 0.001$). It is noticeable, that although the peaks of both densities are close together, the fake headlines tend to be comprised out of more words than the correct ones (which is also reflected by different means: 6.33 vs 5.21). Furthermore, the correct headlines exhibit a lower variance in the number of words.

\begin{figure}
    \centering
    \includegraphics[width=\textwidth]{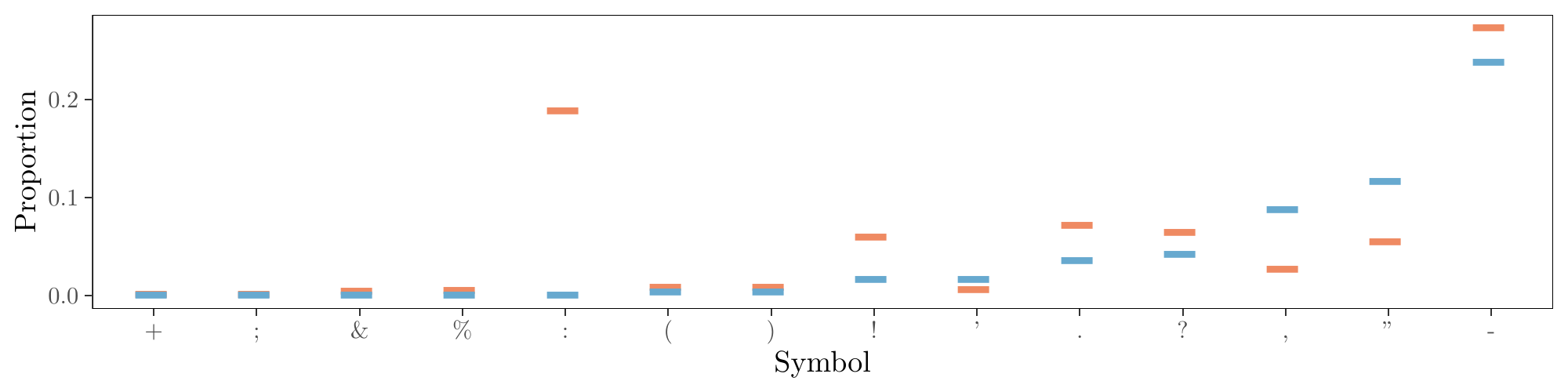}
    \caption{Usage of punctuation marks and special characters in fake and correct headlines.}
    \label{fig:character_analysis}
\end{figure}

In Figure~\ref{fig:character_analysis} the usage of punctuation marks and special characters (x-axis) in correct headlines and fake headlines is depicted. The relative number\footnote{For normalization the number of fake/correct headlines containing the character or punctuation is divided by the total number of fake/correct headlines.} of headlines containing the character or punctuation is displayed on the y-axis. The relative number of correct headlines is represented in blue, and fake headlines in red. The most prominent finding yielded by this Figure is that colons were a striking stylistic device  in fake headlines but never occurred in correct ones.

Even though the differences are much smaller, exclamation marks, question marks, full stops, and hyphens are more frequent in the fake than in the correct headlines. 
On the other hand, the opposite applies to quotation marks, commas, and apostrophes, which occur more often in the correct headlines. 

\begin{figure}
    \centering
    \includegraphics[width=\textwidth]{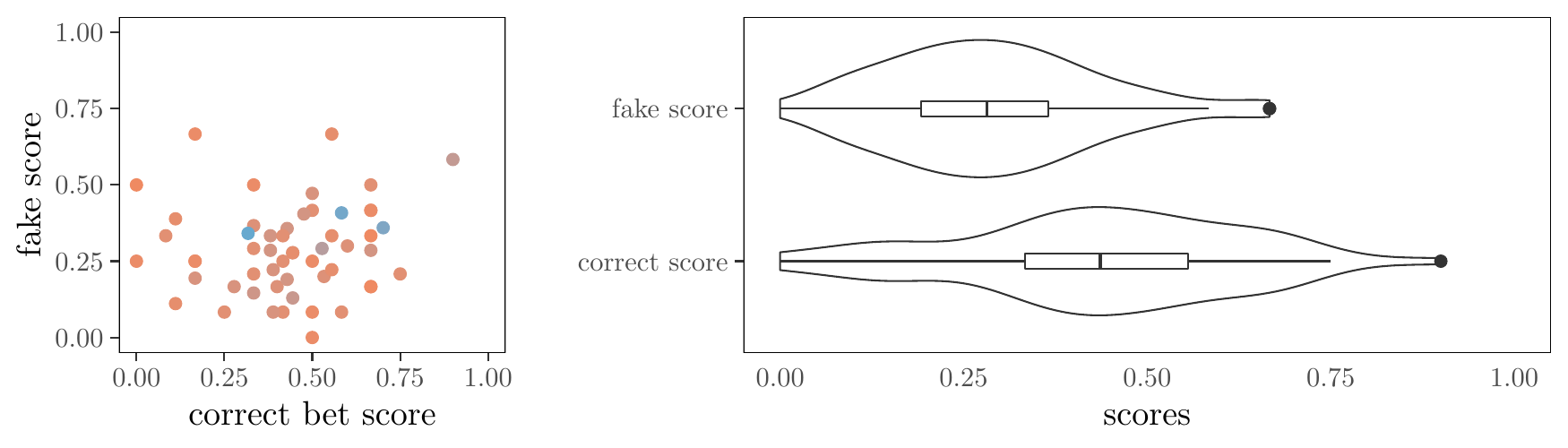}
    \caption{Deviation of fake and correct bet score per user.}
    \label{fig:scores}
\end{figure}

In Figure~\ref{fig:scores} the relative score for fooling and correct bets per player are depicted. For normalization purposes, the total number of points achieved by fooling other players is divided by the number of games times the maximum score\footnote{Fooling two opponents in each of the three rounds sums up in a maximum fake score of 18 (= (3+3)*3).}, which can be achieved in one game by fooling other players. The same is done for the total number of points achieved by betting the correct headline. In this case the number of games is multiplied by the maximum score\footnote{Betting the correct headline three times in a game leads to a maximum correct bet score of 6 (= 2*3).}, which can be achieved by betting three times the right headline.

The left subplot in Figure~\ref{fig:scores} consists of a user scatter plot (fake creation vs. true headline identification). The distributions of the data points, indicate that players differ strongly in their skills. There is no strong correlation between the ability to create good fake headlines and identifying a true headline. While most players are located in the middle area of the scales - meaning, that they received a moderate amount of score points by fooling and correct bets - only a few outliers exist. Outliers at the left upper corner represent players, which are  good at fooling their opponents, but fail more often in finding the correct headline. Outliers at the right upper corner gained the major part of their score points by picking the right headline. The color of the data points indicates the number of games a player completed. The scale reaches from orange (one game) to blue (maximum 25) games. The  number of games is chosen by the individual player. During the evaluation study, participants are allowed to play as much games as they want in a total time range of seven hours. The majority of the participants played between 1 and 6 games. The \textit{super users} of the evaluation study (marked in light blue) are located in the center of the plot, indicating that the relation of their fake score and correct bet score is balanced.

On the right hand side of the Figure, violin plots for the fake and correct bet scores on basis of the individual players are given. Again, score points are normalized by the number of games and the maximum score, which can be achieved. Most of the players chose the right headline in 33 to 56 percent (median = 44 percent) of the rounds.  In contrast to the achieved fake scores, the distribution of points achieved by betting the correct headline is widely dispersed. The values reach from 0 to 0.9, where the latter represents a player who nearly always chose the correct headline. The distribution of the fake scores is more compressed. The majority of players reach relative scores between 19 and 36 percent (median = 28 percent) of the maximum achievable scores for fooling their opponents. The best fake headline creator achieved a relative score of 0.67. 

\begin{figure}
    \centering
    \includegraphics[width=0.9\textwidth]{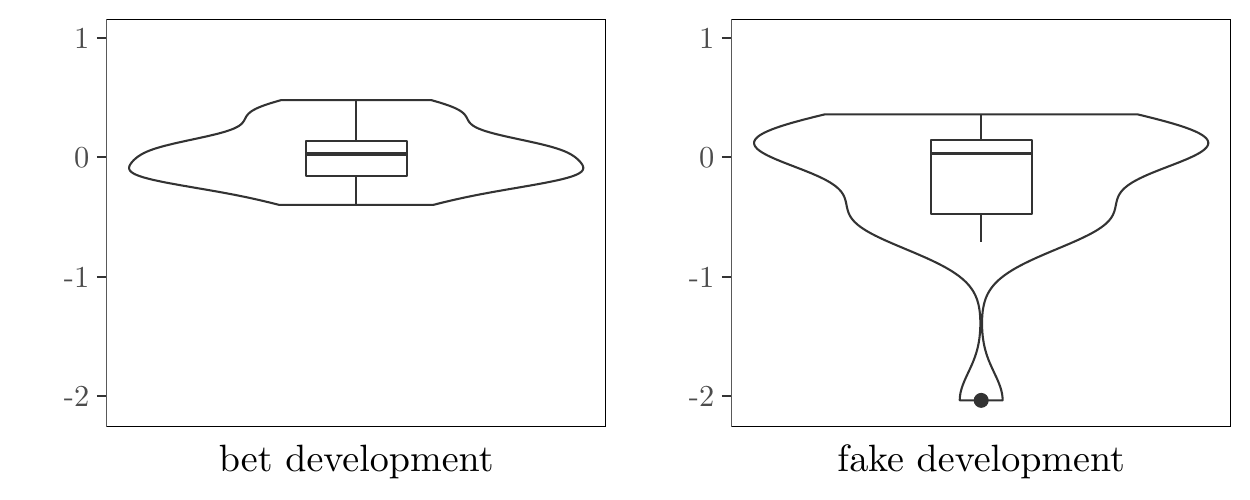}
    \caption{Word frequency density in fake and correct headlines.}
    \label{fig:slope}
\end{figure}

Within Figure~\ref{fig:slope} the temporal development of the players performance in creating convincing fake headlines and betting the correct headline is depicted. To visualize the players learning rate, we first filtered for users, who played at least 16 rounds (which resembles the mean of the sample). The filtering results in 19 participants. 
For each of these participants we fitted a linear model, mapping the number of achieved fake and correct bet score points and played rounds. In a next step, we extracted the slope out of each linear model and compared the values. The comparison of the fake and correct bet developments are visualized within the two boxplots of Figure~\ref{fig:slope}. Negative values indicate a negative trend over time, whereas positive values indicate an improvement of the player. 

For the bet development, neither a downgrade of scores nor a remarkable improvement can be observed. Interestingly the variance within the fake development is  higher. Players tend to get worse in fooling their opponents. Admittedly, the information value of this visualization must be seen critically, as the number of observations is quite small. Further,additional side effects can not be excluded. The game always consists of the two goals "fool opponents" and "bet the correct headline". We do not know, if the ability to chose the right headline might decrease by the fact, that people "learn" to fake, which blurs the results of the performance development.

\subsection{Evaluation of the Gameplay}
As the evaluation study served as a first test for the FakeYou Game application, we asked participants to fill out an online questionnaire to evaluate the game from a user's perspective. Next to age and gender, participants were asked to state how much they liked the game in terms of design and usability. Further, the participants are obliged to report how they perceived their performance and fun level in betting and the creation of fake headlines. Additionally, we asked whether the participant thinks that he/she became better in figuring out the correct headline. 
Two participants thought they got better with every round they played. In the eyes of 15 users new rounds frequently improved their ability to find the correct headline. 18 participants stated that new rounds sometimes raised their awareness towards the wrong headlines. A rare improvement was observed by eight users and only two felt no advancement in their capabilities to identify the fake headlines. Interestingly, the majority of the participants perceived at least a small improvement on their ability to figure out the correct headline. Although this perception is only slightly underpinned by the results reported in Figure~\ref{fig:slope}.

In our study, 31 participants stated that it was always fun to create their own headlines. Furthermore, 14 users frequently enjoyed this process. No one stated that they only sometimes, rarely, or never found joy in the creation of fake headlines.
However, the users suggested further improvements in both comfort options 
as well as bugfixes and server performance.

\section{Discussion \& Conclusion}
\label{sec:discussion}


With this work, we presented a game that is intended to strengthen consumer resilience towards fake news in a gamified setting. Users are pointed to the challenges in detecting fake news and are motivated to think about ways to fake others. The educational effect of both ingredients has to be evaluated further in future work. 
In order to support further evaluation, the game is designed to collect all game and behavioral data of players.
The case study presented in this paper showcased how the game can be applied to get deeper insights into player behavior. Exemplarily, we found for the special case of Spiegel Online headlines and image material that players used different stylistic means for creating headlines. 
Regarding player performance, the comparison of fake and correct bet scores of the players indicated large diversity in game play. The majority of players showed a balanced distribution of fake and correct bet scores. Only a few participants gained their major score points by fooling their opponents with convincing fake headlines. Whereas in sum, the results prefigure that betting the correct headline was easier than fooling other players.

As a typical showcase, our study comes with a few limitations. First of all, only the \textit{Spiegel Online} website was crawled. Certainly, writing styles of headlines differ between newspapers, which might lead to different results in the analysis, but also in the game play itself. However, adjusting the crawler to other websites is straightforward. the crawler can easily be adjusted in order to gather pictures and headlines from other websites. Furthermore, the case study was conducted with only about 50 participants, which were mainly recruited at university. Certainly, a larger and more representative panel of player need to be evaluated in future work.


\section{Acknowlegments}
The research leading to these results received funding by the Federal Ministry of Education and Research, Germany (Project: PropStop, FKZ 16KIS0495K), the federal state of North Rhine-Westphalia and the European Regional Development Fund (EFRE.NRW~2014-2020, Project: MODERAT!, No. CM-2-2-036a), and the Ministry of Culture and Science of the federal state of North Rhine-Westphalia (Project: DemoResil, FKZ 005-1709-0001, EFRE-0801431). All authors appreciate the support of the European Research Center for Information Systems (ERCIS).

%
%
%
\bibliographystyle{splncs04}
\bibliography{literature}

\end{document}